\documentclass[sigconf]{acmart}

\AtBeginDocument{%
  }

\acmYear{2025}\copyrightyear{2025}
\setcopyright{rightsretained}
\acmConference[CSLAW '25]{Symposium on Computer Science and Law}{March 25--27, 2025}{München, Germany}
\acmBooktitle{Symposium on Computer Science and Law (CSLAW '25), March 25--27, 2025, München, Germany}
\acmDOI{10.1145/3709025.3712206}
\acmISBN{979-8-4007-1421-4/25/03}

\usepackage{cleveref}
\usepackage{hyperref}
\usepackage{xurl}
\hypersetup{breaklinks=true}
\usepackage{breakcites}
\usepackage{booktabs}
\usepackage{outlines}
\usepackage{xcolor}
\usepackage{caption}

\begin{document}
\sloppy
\title{When Anti-Fraud Laws Become a Barrier to Computer Science Research}


\author{Madelyne Xiao}
\email{madelyne@princeton.edu}
\affiliation{\institution{Princeton University} \country{USA} \city{Princeton, NJ}}

\author{Andrew Sellars}
\email{sellars@bu.edu}
\affiliation{\institution{Boston University School of Law} \country{USA} \city{Boston, MA}}

\author{Sarah Scheffler}
\email{sscheffl@andrew.cmu.edu}
\affiliation{\institution{Carnegie Mellon University} \country{USA} \city{Pittsburgh, PA}}




\begin{abstract}

Computer science research sometimes brushes with the law,
from red-team exercises that probe the boundaries of authentication mechanisms, to AI research processing copyrighted material, to platform research measuring the behavior of algorithms and users.
U.S.-based computer security research is no stranger to the Computer Fraud and Abuse Act (CFAA) and the Digital Millennium Copyright Act (DMCA)
in a relationship that is still evolving through case law, research practices, changing policies, and legislation.

Amid the landscape computer scientists, lawyers, and policymakers have learned to navigate, \emph{anti-fraud laws} are a surprisingly under-examined challenge for computer science research.
Fraud brings separate issues that are not addressed by the methods for navigating CFAA, DMCA, and Terms of Service that are more familiar in the computer security literature.
Although anti-fraud laws have been discussed to a limited extent in older research on phishing attacks, 
modern computer science researchers are left with little guidance when it comes to navigating issues of  
deception outside the context of pure laboratory research.

In this paper, we analyze and taxonomize the anti-fraud and deception issues that arise in several areas of computer science research.
We find that, despite the lack of attention to these issues in the legal and computer science literature, issues of misrepresented identity or false information that could implicate anti-fraud laws are actually relevant to many methodologies used in computer science research, including penetration testing, web scraping, user studies, sock puppets, social engineering, auditing AI or socio-technical systems, and attacks on artificial intelligence.
We especially highlight the importance of anti-fraud laws in two particular research fields that are of great policy importance in the current moment: attacking or auditing AI systems, and research involving legal identification.

Finally, guided by principles in research ethics, we suggest methods for computer scientists to navigate fraud and identity issues, as well as possible legal paths forward for policymakers to consider.

\end{abstract}

%
%

\maketitle

\section{Introduction}
\label{sec:introduction}

Effective computer science research often requires a degree of impersonation.  Research into increasingly complex systems necessarily requires the research to emulate the behavior of different types of users, or represent to either machines or their operators that they are someone they are not. In so doing, researchers increasingly confront criminal and civil laws that prohibit certain forms of misrepresentation on online platforms. In a United States (U.S.) context,\footnote{As a lawyer and scholars from the United States, we focus our discussion on U.S. law.} this is most broadly in the domain of federal wire fraud (18 U.S.C. §1343), but also state and federal laws concerning identify fraud or identity theft. Research into  AI systems, security systems, social media platforms, or Internet infrastructure can involve deception, and deception can reintroduce some of the criminal legal risk that had been previously relegated. And because some of the most significant research in this domain raises questions about government or political behavior on online systems---e.g., cybersecurity or authentication
for government services, or research into the spread of political disinformation -- governments may possess both means and motive to use criminal law to deter such research.

Many computer science researchers have found themselves in need of legal guidance in the course of their research.
In the United States, this may be most well known in the field of computer security, where researchers sometimes brush with the Computer Fraud and Abuse Act (18 U.S.C. § 1030)  \cite{parkResearchersGuideLegal, mulligan_cybersecurity_2015}, 
the Digital Millennium Copyright Act (17 U.S.C. § 1201; DMCA) \cite{simons_viewpoint_2001,noauthor_thawing_2021},
and perform security analyses that sometimes violate Terms of Service with companies whose products they are analyzing. More recently, social media and other online platforms researchers have experienced similar threats under similar laws, compounded by threats to disable or suspend personal researcher accounts \cite{edelsonUnderstandingEngagementUS2021, longpre_safe_2024}.
Early research into the functions of generative artificial intelligence platforms have begun to receive similar threats. \cite{kumarLegalRisksAdversarial, edwards2024}

Under these commonly-asserted laws, there are now commonly-asserted defenses. While the DMCA prohibits the act of attempting to ``circumvent'' technical tools to access copyright-protected material, there exist both statutory exemptions for reverse engineering (§1201(f)), encryption research (§1201(g)), and security testing (§1201(j)), alongside a more comphrehensive exemption for security research passed as part of the triennial rulemaking conducted by the Library of Congress that modifies the scope of the DMCA (37 C.F.R. § 201.40(b)(1)). In 2021 the United States Supreme Court limited the scope of the CFAA in a useful way in Van Buren v. United States, and the following year the Department of Justice announced a charging policy to avoid charging good-faith security research under the same law. \cite{vanBuren2021, OfficePublicAffairs2022} In more recent litigation, lawyers representing researchers have also found ways to argue against the scope and application of website terms of service, though the full effect of those arguments remains to be seen. 
The reduced application of the CFAA and DMCA has led to a popular belief that most legal risks for software and platform research now are exclusively questions of civil liability.

However, fraud remains relatively under-examined in computer science research.
Although this risk has been mentioned before in research on phishing (see \Cref{sec:related-fraud}), the analysis of how fraud laws apply to existing methods in computer science research is not sufficient for researchers and policymakers to understand the interaction between research and anti-fraud laws. Moreover, as we will discuss in \Cref{sec:methods}, the landscape of computer science research has expanded and changed in such a way that these laws are more relevant than ever before: in particular, the ubiquity and opacity of AI/ML-driven systems will increasingly compel researchers to devise clever and possibly deceptive methods for investigating how these systems work; measuring their impact on platforms and people; and understanding what risks might arise with their continued use. 

In this paper, we study the interactions between end users, researchers, and systems in the context of developments in security research. How can researchers understand and examine these systems in safe, ethical, and legally and technically defensible ways?  

\textbf{Motivation for this work.}
\label{sec:motivation}

This work was inspired by difficulties encountered in the course of planning a study to investigate commercial Internet age verification systems. 
The original study vision included a series of tests intended to determine the effectiveness of simple baseline attacks on these systems: for instance, by having a study participant don simple disguises and use various ID cards to measure the efficacy of the age verification functions. We quickly realized that this line of work would require significant legal and ethical due diligence.

Our research team began designing the study primarily with ethical study design and the CFAA in mind, and worked with a legal clinic to understand the issues. For example, we took significant measures to ensure that our infrastructure never gained ``unauthorized access'' to any system, and we took very strict steps to protect study participants' data. However, once unauthorized access questions were resolved, there were still deceptive elements in the study design: researchers would need to present falsified data to the system to measure its response. 

One recurring theme that emerged from the discussions between the researchers and the legal clinic team was that it was not CFAA that was posing a challenge for our study design.  Rather, it was anti-fraud laws, partly but not entirely arising from our use of government-issued ID documents.
Moreover, as we investigated these questions further, we began to notice that our analyses were
also relevant to significant other swathes of computer security research as well. In particular, the legal and ethical issues were also relevant to most security work that involved deception in investigations of platforms, interactions with ML-driven systems, or interactions with human subjects.

\textbf{Contributions and outline.}
\label{sec:contributions}

Our contributions are as follows:
\begin{enumerate}
\item[\textcolor{black}{(1)}] We define \emph{deception} in the context of computer science research and identify different scenarios of \emph{who}, \emph{by what means}, \emph{to whom}, and \emph{to what ends} (\Cref{sec:methods}).
\item[\textcolor{black}{(2)}] We identify how 
computer science and security research methods engage in various forms of deception (\Cref{sec:taxonomy}).
\item[\textcolor{black}{(3)}] We discuss how deceptive aspects of computer security research relate to fraud in a U.S. context, and situate this new concern in the broader state of legal concerns in computer security
(\Cref{sec:law}).
\item[\textcolor{black}{(4)}] We articulate ethical standards that can guide the future of computer security research that employs deception as a part of its methodology (\Cref{sec:ethics}).
\end{enumerate}

\section{Related Work}
\label{sec:related_work}

The works related to this one largely fall into two main categories: Analyses of computer security or computer science research and their relationship to the law in the U.S., and works that address fraud in computer science research.  We describe each, and we then list other legal concerns we encountered in the literature.

As we described in \Cref{sec:motivation}, there is a fairly robust literature describing the intersection of law and computer science research (especially in computer security and networks).
\cite{parkResearchersGuideLegal} provides the most complete and current review of legal risks for security research.
\cite{longpre_safe_2024} reviews legal and practical dangers to researchers examining AI systems, and calls for both legal protection and mechanisms to ensure technical access by researchers to AI systems.  \cite{kumarLegalRisksAdversarial} also examines legal risks in adversarial machine learning research, a subfield of examining AI systems.
\cite{thomasEthicalIssuesResearch2017} discusses ethical and legal issues arising when researchers use datasets of ``illicit origin'' (e.g. leaked, not necessarily reflecting illicit or deceptive actions by the researchers themselves); they identify many legal issues (computer misuse, copyright, data privacy, use of illicit content like terrorist materials or indecent images, national security issues arising from use of confidential information, and contract violation), but fraud is not among them.
\cite{pfefferkorn2022} reviews how the modern interpretation of the CFAA still remains a vector of concern for cybersecurity research.
\cite{phishing-book} and \cite{soghoian2008legal} concern legal risks of phishing research.
\cite{feltenDigitalMillenniumCopyright2002} discusses the difficulties faced by computer security researchers regarding the Digital Millenium Copyright Act in 2002.
\cite{calo2018} describes the increasingly blurry line between ``hacking'' and ``tricking'' and the intriguing boundaries of the Computer Fraud and Abuse Act.
\cite{hantkeWhereAreRed2024} also provides legal and ethical guidance for network security researchers primarily within the German legal environment.

Another category of related work focuses on the chilling effects that arise in research as a result of legal uncertainty or threat, and discusses the negative consequences to society and individuals arising from that difficulty.
\cite{bhandari2024} connects modern attempts to audit platforms for discrimination with the longstanding auditing done in the offline world and describes negative impact on individuals from lack of access to this research; \cite{baranetsky2018} looks at analogous concerns for data journalists.
\cite{marwick2016} reviews best practices for researchers to avoid harassment related to their work on online platforms.
\cite{mulligan_cybersecurity_2015} is the result of a workshop funded by the National Science Foundation meant to address these challenges.

\label{sec:related-fraud}
Although fraud specifically is not mentioned much in modern computer security research, it does arise in the context of phishing research around 2008.  We highlight three works in particular.

\cite{soghoian2008legal} warns that some ``in-the-wild'' research on phishing, including that of \cite{jakobssonWhyHowPerform2008}, could be considered fraud. He provides best practices for researchers, but the article focuses on other aspects of phishing research that are legally fraught and does not analyze fraud outside the context of the CFAA.

Section 17.5 of \cite{phishing-book} discusses legal considerations in phishing research, including fraud. The analysis relies on a definition of ``defraud'' as meaning ``to take advantage of someone,'' and suggests that since researchers do not have ``intent'' to benefit from the advantage, the case is closed.  We see the legal issue as significantly deeper, as we will discuss in \Cref{sec:law}. \cite{phishing-book} also highlights the difficulty of research involving legal IDs or other ``authentication features,'' which we discuss in \Cref{sec:ids} and \Cref{sec:law}.

\cite{jakobssonWhyHowPerform2008}, entitled ``Why and How to Perform Fraud Experiments,'' discusses performing naturalistic ``in-the-wild'' experiments on fraud compared to laboratory experiments, and describes studies that, in essence, measures users' response to phishing attempts by phishing them (but throwing out the results afterward).  As the work itself points out, ``such experiments pose a thorny ethical issue: if a study is identical to reality, then the study itself constitutes a real-world fraud attempt.''  We will examine ethical issues in \Cref{sec:ethics}.

We also encountered many other U.S. laws and regulations that might impact computer security or computer science research to some extent.  These include
the Electronic Communications Privacy Act (ECPA), trade secret law, export controls, \cite{parkResearchersGuideLegal}
the federal CAN-SPAM Act, the FTC Act, \cite{phishing-book}
trademark law, various state laws \cite{phishing-book,soghoian2008legal}, the federal Wiretap Act \cite{ohmLegalIssuesSurrounding2007}, and
trespass to chattels \cite{soghoian2008legal}.
However, we do not discuss these further in this work.

\section{Uses of deception in security research}
\label{sec:methods}

In this section, we present working definitions of ``deception'' and ``deceptive acts.'' We then introduce a taxonomy of security research methodologies and describe how elements of deception have historically been deployed in each (\Cref{sec:taxonomy}). Following, we present notes on identity verification research and discuss how this line of work presents a unique challenge: it implicates multiple subcategories in our methods taxonomy \textit{and} blurs the line between means and ends (\Cref{sec:id_work}). We note that descriptions and examples provided in this section are intended to be representative, and are not exhaustive surveys of the research landscape.

\subsection{Defining ``deception''}
\label{sec:def-deception}

In the broadest of terms, we define \textit{deception} as \textit{an intentional act of misrepresentation}. 
We separate a ``deceptive act'' into four component parts: the deceiver (``who''), the mode of misrepresentation (``by what means''), the deceived (``whom''), and the purpose of the deception (``to what ends''). In our discussion of the second of these, we also include brief descriptions of what forms of identification or action constitute ``authentic'' or non-deceptive representations in the same setting. 
\newline

\noindent\fbox{%
    \parbox{\linewidth}{ 
    \textit{\textbf{Deception}} in the context of security research occurs when (1) a researcher or system controlled by a researcher (2) makes a false or misleading statement about their identity, their motivations or their intent, or some other aspect of their research methodology to (3) another party or system controlled by a party, in order to (4) obtain information about or access to a system, or to measure the success of, or response to, the deception.}
}
\newline

Here we further explore each of these components (\textit{who} is deceiving \textit{whom}? \textit{By what means}, and \textit{to what ends}?). These real and hypothetical examples are in preparation for the following section (\Cref{sec:taxonomy}) which describes computer security research methods and how they might be viewed as deception.

\subsubsection{Who?} 

\textbf{Human actors}, including researchers and study surrogates acting on behalf of researchers, might employ deceptive techniques in order to understand human and automated responses to the deception in an experimental setting. Employees of companies with in-house or third party red teams might use deception to identify vulnerabilities in an internal computer system. Researchers might enter false information into an AI or other automated system to influence its response. \textbf{Automated systems}, including AI and ML-driven systems deployed by researchers and non-AI/ML driven programs (e.g., automated tenant screening \cite{oyama2009not}), might generate sham data or behavioral nudges in the course of (e.g.) A/B testing or enforcing rate limits. 

\subsubsection{By what means?}

Actors might use \textbf{false names}, or use real names in a misleading way.
They might submit \textbf{false email addresses, phone numbers, or other information} to login processes, or use real email addresses or phone numbers in a misleading way. This includes spoofing (i.e., sending illegitimate messages from) real email addresses or phone numbers and other use of real email addresses or phone numbers that does not involve sending messages from those accounts.

Actors might use \textbf{false account credentials or devices}, or use real account credentials or devices in a misleading way. This includes, but is not limited to, attempting to log in to a computer using compromised credentials and stating the compromised credentials to a person. 
Actors might also use \textbf{false data or spoofed network/device information}, and may also use misleading, altered, or spoofed \textbf{metadata} (e.g. to convince a system they are in a particular physical location).

\textbf{Biometric spoofing} entails use of substitute or fake biological evidence to circumvent biometric verification systems, such as face and fingerprint scanners. Researchers have demonstrated that it is possible to circumvent fingerprint scans by presenting prints lifted from a glass surface using a tacky material (such as the gelatin in a gummy bear) to a scanner \cite{matsumoto2002impact}. \textbf{Fake and modified IDs} can be used (in research and the real world) to evade verification systems that use government-issued identity documents. Depending on jurisdiction and nature of the modification, however, this might constitute a criminal offense \cite{hunt2023kind}. \textbf{Communication of false, misleading, or incomplete information} intersects with many of the previously mentioned modes of deception; this modality additionally includes textual information sent or withheld from online platforms, email communication, and prompts sent to LLMs or other AI models.

\subsubsection{Whom?}

The intended targets of these deceptions (and affected parties, more generally) can include \textbf{human actors}, such as users of an online platform; employees of companies that host web services of interest to researchers; government employees; and participants in a research study. Just as often, \textbf{computer systems} are the direct targets of these deceptions: for instance, hackers might attempt to gain access to login or other credential systems, computer networks, and AI systems. They might adopt any of the approaches mentioned in the previous section in order to spoof credentials, jailbreak guardrails, or overwhelm a distributed network with a 51\% attack \cite{aponte202151}.

\subsubsection{To what ends?}
In our earlier definition of deception, we broadly summarized the objectives of security research with deceptive elements as follows: 
to obtain a) \textbf{information} about or b) \textbf{access} to a system, or 3) to perform \textbf{measurement} of the success of the deception. Among other activities, a) would include most \textit{platform studies}, measurement and/or social engineering experiments that require researchers to perturb an existing social network to understand how human users or the system might respond to those perturbations. The measurement might encompass a combination of human and system-wide responses (e.g., ``how many users have to click on my fake ads before Facebook takes them down?''). b) includes many activities that might qualify as ``traditional security research'' methodologies: e.g., penetration testing and red teaming. c) comprises identity-spoofing red teaming attacks and a number of newer adversarial ML techniques targeting LLMs; the common thread across these works is that the efficacy of the adversarial technique is generally a primary research outcome because the target system is a black box.

\subsection{Deception in practice: a taxonomy}
\label{sec:taxonomy}

We present a taxonomy of methods in computer science research that may implicate fraud. We drew on existing taxonomies and analyses in similar areas in order to formulate this section: \cite{sandvig2014} discusses research methods for investigating algorithmic discrimination on platforms; \cite{gilbert2024risks} reports on legal difficulties in modern behavioral science; \cite{parkResearchersGuideLegal} presents a guide to navigating the legal landscape for security researchers; and
\cite{thomasEthicalIssuesResearch2017} discusses works that involve datasets of illicit origin.
Using the taxonomies set forth in these works as a starting point, we queried Google Scholar for all search queries resulting from the Cartesian product of the following sets: $\texttt{methods} = \lbrace$\textit{penetration testing, red teaming, adversarial ML, reverse engineering, social engineering, sock puppets, undercover studies, user studies, web scraping, scripted studies}$\rbrace$ and $\texttt{axes} = \lbrace$\textit{research ethics, fraud, legality}$\rbrace$. We include social science publications alongside works appearing in the proceedings of tier 1 security venues in 2024.

For each taxon (denoted by italic font), we provide 1) a description of the study methodology; 2) a description of potentially deceptive elements of this research methodology; and 3) reference(s) to academic work employing this method in conjunction with some form of deception. Where relevant, we highlight works whose authors responsibly disclose deceptive elements of their work to participants at the conclusion of the study; we also discuss works where an ethics review or responsible disclosure would ideally be included. See Table \ref{tab:research_taxonomy} for a summary.

Our intention is to emphasize the prevalence and importance of deception in good faith computer science research. 
Note, also, that the descriptions and examples we provide in each taxon are intended to be representative summaries and are non-exhaustive.

\subsubsection{Penetration testing, red teaming, and reverse engineering} 
\label{sec:pentesting}
``Traditional'' security red teaming entails whitehat hacking of a software system in order to understand its vulnerabilities. Oftentimes, the attacker and target are (or are acting on behalf of) the same entity---e.g., a company might employ in-house security red teams in order to test the robustness of its own systems. In particular, red teaming might require an attacker to misrepresent their true identity at the technical level via spoofed network information or incorrect credentials.
\textbf{Example studies:} 
\cite{gallowayPracticalAttacksDNS2024} carries out attacks on Domain Name System reputation systems (i.e., systems that attempt to rate website domains as benign vs. spam vs. phishing, etc). Acting as an attacker controlling a domain, the attacker (researcher) takes actions such as adding unrelated highly-rated benign links to their DNS server to trick the reputation system into giving the attacker (researcher) a higher reputation score, despite the fact that the domain contained binaries that should have been classified as malware (modified by the researchers to remove the actual malware). The researchers conducted this research in concert with a security vendor to ensure the effects were ``isolated,'' but they were interacting with real systems in a deceptive way.
As another example, the first experiment of \cite{maFakeBehalfImperceptibleEmail} involves sending emails from spoofed addresses (e.g. ``admin@google.com'') to several email providers in order to measure how those providers implement various spoofing protections. \cite{maFakeBehalfImperceptibleEmail} disclosed all vulnerabilities to the affected service provider; the authors' study protocol received an exemption from their IRB.

\subsubsection{Web scraping and measurement}
\label{sec:passive_mmt}
Web scraping refers to the practice of using automated tooling to collect information from Internet webpages; snapshots are static captures of Internet webpages (comprising text, image, and metadata contents). In general, social media companies do not make codebases for their ranking and content moderation algorithms publicly available; it is similarly difficult to compel these companies to disclose their data collection practices to the general public.\footnote{The EU has made some progress in this regard, e.g. in the Digital Services Act.} As such, researchers interested in platform accountability work conduct data scrapes, audits, and qualitative analyses of platform content (among other techniques) in order to evaluate platform transparency. 
\textbf{Example studies:} 
\cite{kreibichSpamCampaignTrail2008} ``infiltrates'' a botnet by scraping and reverse engineering the Storm botnet's distribution platform., and then acting as a worker within the net.
\cite{decary2014policing} deploys bots in IRC chat rooms frequented by hackers in order to record chat logs without disclosing the presence of bot accounts to chat room participants. 

\subsubsection{Scripted user studies}
\label{sec:scripted}
Researchers hire study participants to perform a pre-specified sequence of actions (e.g., an information retrieval task) as described by an \textit{activity script}. Participants might be asked to use their own identity information or social media accounts in order to perform these actions. In general, the motivation for these methods (as opposed to sock puppets) is ecological validity: observer effects resulting from inauthentic actions online or new, historyless accounts might invalidate collected data. 
\textbf{Example studies:} 
\cite{dimkov2010two} records two instances of ``physical'' red teaming that require a study subject to use social engineering to obtain ID credentials to gain entry to a protected room.
\cite{haeder2016secret} uses ``secret shoppers'' to evaluate insurance policy holders' ease of access to healthcare providers. \cite{dimkov2010two} disclosed the deception to employees affected by the social engineering experiment; \cite{haeder2016secret} does not mention a disclosure process.

\subsubsection{Sock puppets}
\label{sec:sock_puppets}
Researchers devise false identities in order to perform testing at scale, to access systems they would otherwise be unable to access using their own credentials, or to create controlled user histories for interacting with platform algorithms. In most cases, researchers take pains to make sure that these fabricated identities do not coincide with the identifying information of known living persons. Researchers performing platform studies with sock puppets often deliberately obfuscate the fact that they are operating fake accounts in order to preserve the ecological validity of their results. \textbf{Example studies:} We discuss four studies that use sock puppet accounts to interact with their respective social media platforms. \cite{boshmaf2011socialbot}
deploys Socialbot Network, comprising 102 bot accounts, to connect with more than eight thousand (real) Facebook users. These users did not know that they were interacting with fake accounts. \cite{bandy2021more} creates eight sock puppet accounts on Twitter, manipulate each account's actions, then observe changes in source and topic diversity in each account's timeline. \cite{srba2023auditing} and \cite{hosseinmardi2024causally} create sock puppet accounts on Youtube in order to observe how the platform's recommendations evolve to accommodate artificially curated watch histories. For both \cite{bandy2021more} and \cite{srba2023auditing}, the target of the deception was the recommendation algorithm. Work similar to that of \cite{boshmaf2011socialbot}, in which real world users interact with fake accounts, would ideally receive review from an IRB; after each interaction, or at the conclusion of the study, users on the platform that interacted with a bot account should be debriefed about the true nature of those interactions.  

\subsubsection{Social engineering and interactive studies}
\label{sec:soc_eng}
Researchers interact with web users without informing users of their intention to 1) alter elements of the online environment(s) or 2) observe the effects that these changes have on users. Deception is a central element of this work: by design, the research subject is not privy to their own participation in experimental work and, as a result, cannot give informed consent. \textbf{Example studies:} \cite{di2022revisiting} conducts a longitudinal analysis of web administrators' responses to individual user requests for personal information; the requests were sent by the research team under aliases and included modified ID information (see \Cref{sec:id_work}). The study authors received consent from the owners of the credentials used to submit data requests. At the conclusion of the study, the study authors contacted the organizations targeted by their data requests to disclose their study objectives and inform them of their findings. 
\cite{acharyaConningCryptoConman2024} created tweets ``tailored to lure scammers,''  ``bait [scammers] into revealing their payment information,'' and then validate the scam by ``observ[ing] private key theft by scammers.''  These straightforwardly interacted with scammers in the wild in a deceptive way.
Additionally, much phishing research falls into this category: \cite{jakobssonWhyHowPerform2008} describes experimental designs employing deceptive URLs and email addresses; more recently, \cite{royChatbotsPhishbotsPhishing2024} investigates similar questions, but includes LLM-generated phishing attacks.

\subsubsection{Audits} 
\label{sec:audits}
Most social media platforms (and other online services) do not make code and data readily available to the public. Researchers employ a variety of techniques in order to infer under-hood mechanisms for platform operations (e.g., content moderation and ranking): for instance, counterfactual testing, robustness and sensitivity testing, and explainability analyses. 
\textbf{Example studies:}
\cite{kaplanMeasurementAnalysisImplied2022} measures discrimination in Facebook ad distribution by posing as a Facebook advertiser and running job ads to determine potential biases in their distribution. The authors took measures to minimize harm to Facebook users by only creating ads for real job listings. These ads only displayed text germane to the actual job listing, and no personal information was collected about users who clicked on each fake ad. 
\cite{westPictureWorth5002024} analyzes ML tasks in mobile apps; they detected the tasks, reconstructed the pipeline, and then assessed performance.  The performance assessment portion involved a script that would inject data inputs into the model and measure results. 
As AI models have gotten more effective at doing voice and face recognition, researchers evaluating those systems present false data to them in the process of evaluating them (e.g. \cite{jiangCanHearYour,chengALIFLowCostAdversarial2024,kimScoresTellEverything2024}).

\subsubsection{Attacks on machine learning and artificial intelligence} 
\label{sec:adversarial_ml}
In order to understand the robustness of ML-driven systems, researchers use prompt engineering (in the case of LLMs), poisoning attacks, and jailbreaking techniques to identify vulnerabilities at all points in the ML pipeline: training set contents, model weights, and the scope of the reported evaluation. Some of these approaches might require researchers to misrepresent themselves to the system under analysis. For example, researchers might prompt an LLM as though they were someone else to cause a change in the response or to reverse-engineer the model or its training data. In recent times, image forgery attacks have been aided by advanced image generation networks \cite{cox2024}. Researchers might also measure the susceptibility of a model to data poisoning by attempting to cause the model to train on false or misleading data. 

Additionally, researchers might probe the boundaries of watermarks, guardrails, or other restrictions on the outputs of generative AI models (see, for instance, the ``Grandma jailbreak'' for LLMs \cite{zhang2023jailbreak}); toward this end, researchers might feed the AI false information. In the coming years, we expect that attacks, defenses, and models in this space will evolve rapidly and additional types of attacks will become relevant---including those that contain false representations. As well, we expect that research in these fields will become even more important as these systems are used for more social and technical purposes. \cite{carliniAreAlignedNeural} is a recent example of such an attack on ``live'' large language models.

We do not provide an exhaustive taxonomy of ML attacks in this section; see references for a discussion of research on intelligent systems, and see \cite{kumarLegalRisksAdversarial} for a discussion on the CFAA implications of adversarial ML research.
\textbf{Example studies:}
\cite{perez2022red} uses language models to red team LLMs.
\cite{yanLLMAssistedEasyTriggerBackdoor} attacks code completion models to inject vulnerabilities by poisoning the model's data and causing it to yield code in which vulnerabilities are hard to detect. \cite{carta2022video} and \cite{scherhag2017biometric} use video stream injection and facial morphing applications, respectively, to circumvent biometric face scans. 
Several systems investigate jailbreaking of models, often by crafting prompts in such a way that harmful instructions are ``disguised'' to avoid guardrails (e.g. \cite{carliniAreAlignedNeural,liuMakingThemAsk2024,yangSneakyPromptJailbreakingTextimage2024,yuDontListenMe2024}).
Works that generate deepfakes (for example, to build databases of deepfakes for other researchers to use) also mention the need to pay special attention to legal consideration since this is a developing area (e.g. \cite{mengAVAInconspicuousAttribute2024,laytonSokGoodBad}).
Although LLMs are much more widespread now, this line of research to understand ``adversarial'' inputs causing unexpected results in machine learning models started much earlier \cite{huangAdversarialMachineLearning,kurakinAdversarialMachineLearning2017}.

\subsection{Research involving legal identification}
\label{sec:id_work}
\label{sec:ids}
The age verification study we describe in~\Cref{sec:motivation} implicates several of these approaches, including red teaming and auditing.
However, that work and others like it (for instance, \cite{di2022revisiting, di2019personal}) raise additional questions because they require that researchers present or alter identity documents.
To the extent that these images are significantly different from their own identities, these methods constitute a form of deception, undertaken in the service of understanding the mechanism and performance of these verification algorithms. In \Cref{sec:law}, we discuss why research involving modification of legal ID documents is uniquely legally challenging.

\cite{guanIDNetNovelDataset2024} and \cite{lerougeDocXPand25kLargeDiverse2024} both build datasets dataset of synthetic ID cards from a training dataset of 5.8k screenshots of real IDs and photos. Xie et al.\ additionally create fraudulent versions of the synthetic IDs for further training.

\section{Legal and policy considerations}
\label{sec:legal}
\label{sec:law}

In this section we review how the increased use of deception in research can translate to legal risk. As has been long discussed, and summarized most recently by \cite{parkResearchersGuideLegal}, researchers conducting studies into software systems and platforms must routinely consider how their activity may face civil or criminal threats.

Before turning to specific laws and their application, it is important to emphasize how uncertainty can affect academic researchers in ways unlike other actors. Commercial actors also experience some of the concerns, such as those related to web scraping. But those actors are comparatively better equipped to take risks around liability. This is perhaps best evidenced by the litany of artificial intelligence companies that built their systems on scraped data, only to now spend the next several years litigating how lawful that was \cite{EthicalTech2023} .

Researchers can rarely take that risk. Beyond their comparatively fewer resources, the consequences of having work being labeled "unlawful" can provide a deeper chill. Universities and other institutions may pressure researchers to back off of work, lest the institution become the "test case" for a new theory of liability \cite{parkResearchersGuideLegal}. Funders may cut support out of concerns of follow-on liability \cite{gilbert2024risks}. Publishers may be disinclined to publish work if the work is deemed unlawful, even if the claim is dubious \cite{mulligan_cybersecurity_2015}. In short, as Bhandari recently summarized, "[f]or every researcher or journalist willing to conduct important research in the face of such uncertainty, there are likely untold numbers of others who would engage in such work but cannot in light of the risks attached" \cite{bhandari2024}.

\subsection{The rise of fraud}
\label{sec:rise_of_fraud}
Classic legal concerns for software and platform research in the United States usually amount to a combination of the Computer Fraud and Abuse Act (CFAA), the Digital Millennium Copyright Act (DMCA), and concerns about claims rooted in breaches of terms of service. Years of litigation and policy debates have gradually developed some degree of consistency or at least common lines of argumentation within those domains, which are reviewed in \Cref{sec:cfaa-dmca}.

An increased reliance on deception as a research tactic has put a focus on an entirely different domain, for which there has been little to no line drawing: claims rooted in fraud.\footnote{This article considers fraud as a standalone legal cause of action, as opposed how fraud is used as an aggravating factor in other laws or as a defense in certain civil claims \cite{podgor_criminal_1999}. It also does not address the ways in which statutes can sound in fraud but be applied in broader ways, such as the law against a conspiracy to "defraud the United States," which includes monetary fraud where the United States is the victim, alongside broader notions of obstruction of government action. 18 U.S.C. § 371.} There is no single statute that encompasses all of the concept of "fraud." It is instead a collection of civil and criminal statutes that all generally operate around the notion that when a person deceives a person and thereby deprives them of something of value, that should be the basis of liability. The branches of modern fraud law owe their origin to the common law tort of deceit, established well over a century prior and with roots that stretch back considerably further \cite{henricksen2021}. Yet despite their history, the scope of these laws is notoriously unclear and constantly evolving \cite{podgor_criminal_1999}.

While there are countless state and federal laws that can be brought against alleged fraud, the most relevant at the federal level are the federal wire fraud statute (18 U.S.C. § 1343), the identity fraud and theft statutes (18 U.S.C. §§ 1028 and 1028A), the "access device" fraud statute (18 U.S.C. § 1029), and the fraud-specific provisions of the CFAA (18 U.S.C. § 1030(a)(4)). Each has unique aspects, but the most capacious is wire fraud, which requires a prosecutor to show that the defendant made (1) a material deception; (2) with an intent to defraud; (3) through the use of interstate wire, radio, or television communication; (4) which, if successful, would have resulted in a loss of money or property \cite{desantis2018}. Its most famous application to research activity was likely the prosecution of Aaron Swartz, where it was asserted alongside the CFAA on the theory that Swartz's scraping of the JSTOR database was wire fraud due to the efforts he undertook to conceal his identity and appear as other web traffic \cite{sellars_impact_2013}. At the federal level most of these require some sort of nexus to interstate communication or commerce, but that has not proved to be a meaningful limit on claims for online activity. Some courts infer interstate communications by the absence of a relevant server in the parties' state, and other courts allow the mere fact that the communication was sent over the Internet to suffice to show requisite interstate wire communication \cite{malkiel2023}.

The rise of risk in fraud also re-introduces the government as a potential threat actor. With the ebb of the CFAA and DMCA reviewed below, and the lack of criminal enforcement of terms of service generally, the risk of \textit{criminal} prosecution for platform research has reduced considerably over time. Most of the concern has shifted to civil actions by the platforms themselves. This is not to say the government has not been a more general threat; in fact, the past few years have underscored just how much government actors can deter unwanted research through congressional subpoenas, cuts to or conditions on funding, and attempts by the government to more directly control the actions of universities \cite{calo_american_2024}. But the rise of claims of fraud would provide even more government actors further ways to surveil, pressure, and otherwise coerce the actions of research in the name of law enforcement. For example, a state seeking to impose mandatory age verification on certain websites might be tempted to use criminal law to target researchers and advocates who question the legitimacy of age estimation tools. A similar situation arose in 2021, when the Governor of Missouri threatened criminal action against a researcher who discovered and disclosed how a government website leaked private data \cite{brodkin_missouri_2021}. Attorneys general and prosecutors critical of social media misinformation research (which is often perceived by conservatives as a threat against their political speech) risk finding new ways to exert pressure on such research when it involves some form of deception. Criminal fraud laws have long been viewed as a substantial risk for prosecutorial overreach \cite{podgor_criminal_1999}. The rise of deception-based research in areas where the government actors may have strong feelings provides both means and motive to threaten such research.

But past fraud litigation against another public interest group---undercover journalists---may provide some insights on how courts can effectively balance claims of fraud against socially valuable newsgathering. Courts have not been consistent in how they approach claims against undercover journalists,\footnote{\textit{Compare} Dietemann v. Time, Inc., 449 F.2d 245 (9th Cir. 1971) (upholding a breach of privacy claim against an undercover journalist who brought a hidden camera into the home office of dubious homeopathic therapies), \textit{with} Desnick v. ABC, Inc., 44 F.3d 1345 (7th Cir. 1995) (denying claims for trespass, wiretapping, fraud, and breach of privacy against undercover journalist who brought a hidden camera into ophthalmologist's office). \textit{Desnick} itself distinguished \textit{Dietemann} by drawing a distinction between use of fraud to enter a business versus use of fraud to enter a home office. \textit{Desnick}, 44 F.3d at 1352–53.} but have provided useful instruction in how to consider their liability against possible claims of fraud which will be relevant to deception research as well.

The following subsections will chart out some common questions or issues in fraud litigation, how they may apply to the research reviewed above, and what past cases against journalists can provide as insights on how to reconcile the inherent tensions present.

\subsubsection{Information as property}
While not all fraud laws explicitly require this \cite{podgor_criminal_1999}, for federal wire fraud one must show that the defendant's scheme worked to deprive (or, if successful, would have deprived) its victim of money or property. The Supreme Court recently reaffirmed that this is tied to "traditional property interests" of the victim, and not a broader set of informational or dignitary harms.\footnote{Ciminelli v. United States, 143 S. Ct. 1121, 1128 (2023).} But in Carpenter v. United States, a case concerning a journalist who leaked news stories to investors ahead of publication, the Court has also said that confidential business information can be the sort of property that fits that definition.\footnote{Carpenter v. United States, 484 U.S. 19, 25 (1987); \textit{see also} United States v. Abouammo, No. 19-cr-621, 2021 WL 718842 at *4 (N.D. Cal. Feb. 24, 2021) (Twitter's confidential user account information was "property" for purposes of § 1343).} It is therefore insufficient for a researcher to evade liability by arguing they did not benefit monetarily. The fact that the researcher learned company secrets through the deception may keep wire fraud and related laws in play. In this manner, the propertization of company secrets presents a similar danger as it does for efforts to mandate algorithmic transparency through legislation or regulation. As reviewed by \cite{kapczynski2020}, the recognition of confidential business information as property in \textit{Carpenter} and in the Takings Clause case Ruckelshaus v. Monsanto Co.\footnote{467 U.S. 986 (1984).} allows companies to challenge mandated transparency by arguing it amounts to a taking to which they are entitled compensation.

What may save deception-based research from these sorts of claims is an insight from cases examining fraud in undercover reporting. In Desnick v. ABC, Inc, a federal appeals court rejected a theory of civil fraud asserted against a reporter in part because "[t]he only scheme here was a scheme to expose publicly any bad practices the investigative team discovered, and that is not a fraudulent scheme."\footnote{\textit{Desnick}, 44 F.3d at 1355.} In other words, if the information the deception uncovers shows that the company is engaged in some sort of inappropriate behavior, some courts appear less likely to uphold a fraud claim for obtaining that information.

\subsubsection{Deception and materiality}
Further complications arise around the "how" and "to whom" of deception. Within wire fraud, the Supreme Court has already established that false or misleading representation must be \textit{material} to the victim,\footnote{Neder v. United States, 527 U.S. 1 (1999).} Civil claims of fraud often instead emphasize not just materiality, but \textit{reliance} \cite{goldberg_place_2006}. That is, that the tort plaintiff in a civil lawsuit must show that they personally depended upon the false or misleading statement to their detriment. Different forms of platform research will perform differently under these distinct tests, but it is certainly possible that deception done more for the purposes of evading detection of ongoing research, as opposed to a direct representation to another, may fare better under at least the civil formulation. On the outer edges of materiality, courts split on whether a victim who gets exactly what they hoped to out of a transaction is actionably defrauded if they would not have initiated the transaction but for the deception \cite{frohock2020}.  For research where a researcher acts as a customer on a website and indeed completes a transaction, this difference in approaches may be significant.

A noteworthy distinction between deception in platform research and most forms of conventional fraud is that very often the object deceived is not a person at all. It is a computational system. \cite{calo2018} considers the question of whether "tricking" a computational system should be thought of as hacking under the CFAA, but their analysis does not examine whether this is actually \textit{fraud} under the relevant provision of the CFAA (§ 1030(a)(4)), instead focusing on the \textit{damage} provisions of §1030(a)(5). Deceiving a computer was core to the arguments in the Aaron Swartz case, where the prosecution alleged that the techniques Swartz used to download files from JSTOR---including switching his MAC and IP addresses and using IP addresses assigned from MIT---was actionable deception under wire fraud, despite the fact that IP addresses are dynamically assigned and nothing technically requires a computer to keep a consistent MAC address \cite{sellars_impact_2013}.\footnote{A similar point was made by X against a commercial scraper, and rejected by the court that considered it. Meta Platforms, Inc. v. Bright Data Ltd., No. 23-cv-77, 2024 WL 251406 at *7 (N.D. Cal. Jan. 23, 2024) (use of IP proxies not actionable under California's unfair business acts law as an unlawful deception, because IP addresses are inherently dynamic).}

\subsubsection{Artifacts of fraud}
Laws like the federal identity fraud and identity theft statues (18 U.S.C. §§ 1028, 1028A) and the access device fraud statute (18 U.S.C. § 1029) regulate the production, use, and trafficking in specific forms of fraudulent documentation--fake IDs, passports, credit card numbers, serial numbers, and so on. The natural question arises how one can design a study to test how good tools are at detecting inauthentic access documents, if making an artifact that is too successful as a forgery is itself unlawful. How, then, can one conduct such research?

The clues may lie not in undercover journalism,\footnote{Indeed, use of fake IDs for uncover research has led to liability. \textit{See} Planned Parenthood Fed. of Am., Inc. v. Ctr. for Medical Progress, 402 F. Supp. 3d 615 (N.D. Cal. 2019) (allowing a RICO claim to proceed with § 1028 as a predicate act, for an advocacy organization who used a fake ID to entered a Planned Parenthood conference).} but in the world of film and television. Prop makers for films also have no general exemption from these sorts of laws, but regularly make convincing replicas for purposes of film and television. The key in that environment has been a combination of careful tailoring for the use case and effort to mitigate any possible uses of the item that would lead to the social harms these laws are meant to address \cite{geaghan-breiner_how_2022}. In effect, the goal for a prop maker is to create an artifact that is convincing \textit{solely and specifically} for the shot, but wholly unconvincing in other contexts. Collection and destruction of the items after the study can also serve as a practical mitigation. Exercising additional due diligence to ensure that the names used do not correlate to real persons (another common step in film and television production) can also help to mitigate these concerns.

\subsubsection{The First Amendment and Fraud}
Of course, researchers are engaging in their deception for the purpose of producing some sort of academic expression, instead of for financial gain like the typical fraud defendant. Does the First Amendment provide them any help? The short answer is the lawyerly "it depends." Advocates have helped courts to recognize that this form of research has First Amendment value and have protected both its methods of information gathering and dissemination, but what that exactly means is more complicated than one might initially believe.

Prior advocates have helped establish that the First Amendment does in fact extend to information gathering activity done in the course of academic research, and the publication of material like source code in order to illustrate aspects of research \cite{parkResearchersGuideLegal}. Most critically, a court considering a constitutional challenge to the CFAA brought on behalf of academic researchers and a media outlet found that "plaintiffs have a First Amendment interest in harmlessly misrepresenting their identities to target websites."\footnote{Sandvig v. Sessions, 315 F. Supp. 3d 1, 16 (D.D.C. 2018). The court eventually dismissed the case, finding that the CFAA did not reach this activity. Sandvig v. Barr, 451 F. Supp. 3d 73, 92 (D.D.C. 2020).} Cases have also found a First Amendment interest in scraping websites for purposes of research and advocacy.\footnote{\textit{See Sandvig}, 315 F. Supp. 3d at 15; S.C. State Conf. of NAACP v. Kohn, No. 22-cv-1007, 2023 WL 144447 at *6–7 (D.S.C. Jan. 10, 2023) (scraping of government records that have been historically made public is protected by the First Amendment).} This benefits academics and journalists alike \cite{baranetsky2018}. On the publication side, scholars have explained why "crime-facilitating speech," or speech that allows others to understand how to commit certain crimes has public value and should be protected in many contexts, even though it could be put to nefarious use \cite{volokhCrimeFacilitatingSpeech2005, matwyshyn2013}. 

In terms of false speech specifically, the Supreme Court's 2012 decision in United States v. Alvarez\footnote{567 U.S. 709 (2012) (plurality opinion).} is instructive. There, the court struck a "stolen valor" statute that punished those who falsely claimed to receive certain military honors. Its reasoning was rooted in the history by which the Supreme Court approached content-based laws of this nature. Absent a specifically enumerated list "historical and traditional" categories where the state may punish speech, any attempt to control even factually false speech must satisfy strict scrutiny.\footnote{\textit{Id.} at 718.} But, in so doing it upheld past cases that denied First Amendment protection for speech used in certain frauds, and repeatedly stated that speech used in a fraud remained unprotected.\footnote{\textit{Id.} at 719 (citing Illinois ex rel. Madigan v. Telemarketing Assocs., Inc., 538 U.S. 600 (2003)).} What that means precisely remains subject to debate. For its part, the prior case that \textit{Alvarez} cites to make this point emphasizes that "[s]imply labeling an action one for `fraud,' of course, will not carry the day," but that intentional misleading for the purpose of receiving monetary donations would still be actionable.\footnote{\textit{Madigan}, 538 U.S. at 617.} The United States Court of Appeals for the Ninth Circuit began to reconcile this tension in a constitutional challenge brought against Idaho's "Ag-Gag" law, which prohibited use of misrepresentations to enter a food production facility, or to obtain records or employment from a facility. The court in that case used \textit{Alvarez} to strike the provisions that criminalized deception to enter a facility, but not those that proscribed deception to secure employment.\footnote{Animal Legal Defense Fund v. Wasden, 878 F.3d 1184, 1190 (9th Cir. 2018).} 

But even if research methods that use deception are protected by the First Amendment, what that protection actually \textit{means} remains ambiguous. Courts are quick to observe that engaging in journalism or other protected activity does not excuse generally applicable laws that are unrelated to the newsgathering activity.\footnote{\textit{See, e.g.}, \textit{Desnick}, 44 F.3d at 1351 ("[T]here is no journalists' privilege to trespass.")} But for laws that attempt to punish certain types of research activity, however, this allows courts to critically examine whether those laws affect a free speech interest, and if they do, whether they meet requisite scrutiny. For an example outside of the deception context, in 2010 the United States Court of Appeals for the Fourth Circuit considered a free speech challenge to a Virginia law that prohibited the publication of another's Social Security Number. The challenge was brought by an advocate who published the Social Security Numbers of several prominent residents that she obtained from Virginia's public land records, to illustrate the state's poor data handling practices. The court held that punishing the advocate under that law would be unconstitutional.\footnote{Ostergren v. Cuccinelli, 615 F.3d 263, 271 (4th Cir. 2010) (noting that posting the numbers themselves was "integral to her message. Indeed, they \textit{are} her message" (emphasis in original)).} 

Recognizing a free speech interest can help a researcher in other ways as well. It may help inform challenges against government actors, including those that attempt to use pretextual allegations of criminality to chill research activity. Free speech principles animate defenses rooted in "anti-SLAPP" law, or statutes designed to achieve both quick dismissal and attorney's fees for defendants when they receive a legal threat related to some form of public advocacy or petitioning.\footnote{\textit{See, e.g.}, X Corp. v. Ctr. for Countering Digital Hate, Inc., 724 F. Supp. 3d 948, 987 (N.D. Cal. 2024), \textit{appeal docketed}, No. 24-2643 (9th Cir. April 25, 2024) (granting a motion to strike under the California anti-SLAPP law against X's attempt to chill research onto its platform).} It also helps courts push back against attempts to recover damages for reputational harms in non-speech-related tort claims, under a line of cases extending back to the Supreme Court's decision in Hustler Magazine, Inc. v. Falwell.\footnote{485 U.S. 46 (1988); \textit{see, e.g.}, \textit{Ctr. for Countering Digital Hate}, 724 F. Supp. 3d at 976–77 (applying this line of cases to deny recovery of reputational damages against a research-based scraper).} This critical attention to which harms are recoverable has been useful in limiting the overall risk to research that breaches a website's terms of service, as typically the amount of recoverable damages in those cases (assuming the researcher did not seriously damage the website in the process) is at or near zero.

\subsection{Long-discussed legal concerns}
\Cref{sec:cfaa-dmca} describes how computer security resesarch has previously come to work in the context of the CFAA, DMCA, and potential terms of service violations. 

The net effect is a mix for researchers. While these claims remain a source of concern and can chill research in the ways reviewed at the beginning of this section, advocates have found ways around these obstacles that can allow research to proceed.

\subsection{Navigating legal barriers}
Despite considerable barriers to their work, academics conducting security, platform auditing, and adversarial ML studies have found ways to continue their research. Many academics simply do not consider the legal risk, or assume that the ethical analysis conducted by an Institutional Review Board (IRB) also handles any legal concerns. Academic researchers tend to place ethical considerations above legal ones when performing risk analyses \cite{gilbert2024risks}. But a more appropriate path is to consider law and ethics each as important but distinct inquiries, and work to craft a study that addresses both concerns. To that end, we provide here some practical guidance to consider when conducting research that uses deception. Of course, anti-fraud laws are only one of several areas of potential legal risks, and the law can vary considerably in different jurisdictions. When doing deception-based work one should consult with a lawyer that with expertise in computer science research who can help further. \cite{parkResearchersGuideLegal} provides further information about engaging with an attorney for research.

\subsubsection{Appreciate that research involves risk.}
Researchers in this area may need to borrow a mentality from the world of journalism, which has long understood that provocative work in their space inherently involves some risk. The only true risk-free path for a newspaper is not to publish at all. Lawyers, similarly, should think of their role not as risk elimination, but rather risk management. A lawyer is bound by their own rules of ethics not to counsel a client to engage in activity that they know to be criminal or fraudulent,\footnote{ABA Model Rule of Professional Conduct 1.2(d)} but they may provide a candid assessment as to the scope of those laws. And outside of crimes and frauds, a lawyer is permitted to advise a client to, for example, engage in "efficient breach" of a contract when the consequences of doing so are favorable to performing as contracted. This is most relevant here when considering whether a researcher should breach a terms of service, which almost always prohibits any attempt to use the website for something other than the proposed business transaction. A lawyer and their research client might rightly consider the fact that 1) the terms may be unenforceable, 2) companies are unlikely or unwilling to enforce them, or 3) the value of the expected research outcome is worth the potential consequence of enforcement. 

It is also important to remember that different people experience risk differently. Researchers may be more or less able to take a risk depending on their identity and status. A tenured professor stands in very different shoes than a graduate student attending school on a student visa, who may risk having their visa revoked if found guilty of a fraud.

\subsubsection{Consider whether permission is feasible.}
Most of the legal risks under fraud, as well as those under the CFAA and other research-related laws, can be neutralized with the consent of the research target. When the work is not inherently adversarial (or when the platform or vendor welcomes adversarial research as a way for them to test their safeguards) a researcher may be able to obtain consent. Such consent should of course by carefully documented, and use of deception to \textit{obtain} the consent will likely lead to more trouble. Working directly with a vendor may also risk having the vendor steer or influence the research in inappropriate ways, and researchers should hold firm to their independence in such negotiations.

Permission can also be indirect. Depending on the research, the researcher may be able to take advantage of a "bug bounty" or similar open call by a platform or vendor to engage in certain types of research. The researcher should be aware, however, that those bounties will likely come with conditions that may frustrate the goals of the research, including non-disclosure obligations or long embargo windows. Violating the terms of the bounty will make it considerably more difficult to rely on the bounty's permission for the activity.

\subsubsection{Analyze the law carefully and avoid unnecessary deception.}
Careful analysis of the relevant law may also reveal paths forward. In the same vein as our own ID study, \cite{di2019personal} presents modified identification documents to online service providers in order to request personal data collected by the service. Di Martino et al. took pains to ensure that their modifications would not run afoul of anti-forgery and identity theft statutes. The authors found that sensitive information, including handwritten signatures and dates of expiry, could be censored in cases where ID images were required for authentication. As such, they took care to alter only those fields when generating modified ID documents. Additionally, modifications were made to \textit{photocopies} of ID documents only. Similar strategies can help when using the law as a research tool. For example, \cite{borradaile2020whose} used public records law to study social media monitoring done by the Oregon State Police. Knowing that records requests for source code were unlikely to be successful, they instead requested access to code \textit{logs}, which they used for their analysis. 

One should also limit any deception to solely what is needed in order to conduct the study. Borrowing from the world of film and television, any objects that are created in order to further the deception should be tailored to work solely for the research and for no other purpose. Use of one-sided IDs, items of a strikingly different weight than an authentic version, or items with obviously fictitious information can be an effective balance between utility for a study and ineffectiveness in other contexts. But one should be mindful that not all laws or courts agree that a fraudulent item needs to be convincing for a person to be liable.\footnote{\textit{Compare Planned Parenthood}, 402 F. Supp. 3d at 650 (declining to adopt a "sufficient quality" element in a charge for identity fraud under § 1028), \textit{with} United States v. Gomes, 969 F.2d 1290, 1293 (1st Cir. 1992) (a charge for possession of a counterfeit Social Security card could only be met if the counterfeit "possesses enough verisimilitude to deceive an ordinary person")} 

\subsubsection{Avoid obtaining things of value}
Finally, because many laws turn on whether the deception was used to obtain something of value (including, in some cases, confidential information), a researcher should try to come up with ways that cut off any benefit for the deception. The use of sandboxed environments for research are helpful; if the researcher is able to construct the environment such that the \textit{sole} information that is received by the researcher is whether the system \textit{would have} let you in if connected to a live website or system, the researcher can cut off an argument that they actually obtained anything of value.\footnote{It is also possible that access to confidential information that is \textit{never put to use} by the researcher would not amount to wire fraud, but this would be practically challenging to prove and may not be found in all jurisdictions. \textit{See} United States v. Czubinski, 106 F.3d 1069, 1075–76 (1st Cir. 1997) (browsing confidential information out of curiosity, but not using it in any way, does not amount to a scheme to defraud).} Along a similar line, a researcher should budget to pay the customary price for access to any paid system, and not use deception to avoid payment.

\section{Research Ethics for Identity and Fraud}
\label{sec:ethics}

In this section we examine fraud-related ethical considerations in computer science research, aided by but not fully addressed by existing research ethics frameworks.
As Albert and Grimmelmann recently cautioned, one should not equate law with ethics: ``[l]aws are not always right, and they are never neutral,'' \cite{albert_right_2023}; similarly, Thomas et al.\ remind us that ``[o]ccasionally research may be illegal but still ethical,'' \cite{thomasEthicalIssuesResearch2017}. Ethics and law are rightly separate, albeit related, inquiries.
Many jurisdictions require by law that scientific work funded by federal agencies be subject to ethical review (e.g. the ``Common Rule'' \cite{protectionsohrpFederalPolicyProtection2009} in the U.S.).

The formal guidance that computer scientists working with IRBs would be familiar with provides only high-level principles relevant to addressing ethics and especially those related to deception. Similarly although most computer security publication venues require an ethics analysis when relevant \cite{finnEthicsGovernanceDevelopment2023}, in practice they supply little practical advice when considering questions of deception or fraud.
Newer discussions about research ethics specific to computer security are ongoing (e.g. \cite{FosteringResponsibleComputing,cranorConferenceSubmissionReview,kohnoEthicalFrameworksComputer}).

In this section we discuss guidance for tackling these questions in significantly more detail, though each study will pose case-by-case questions that we cannot address fully in an abstract analysis.
This ethical analysis dovetails with the legal analysis both to illuminate guidelines for scientists undertaking research where fraud may be relevant, and also suggests potential paths that might change on the legal side -- in other words, this ethical analysis helps us ensure that our legal analysis works with, not against, ethical considerations.

\textbf{The Menlo Report and its limits.} For computer science research, the typical ethical analysis for protecting human subjects in studies reviewed by an IRB is summarized in the 2012 Menlo Report \cite{menloReport}, similar to the 1979 Belmont Report \cite{belmontReport} for protecting human subjects in biomedical and behavioral research.

The Menlo report centers four lenses with which to consider the ethics of a particular study:
\begin{enumerate}
\item \emph{Respect for Persons}: A rights-based deontological approach that focuses on respecting study participants' and other individuals' rights as autonomous agents (especially via informed consent).  
\item \emph{Beneficence}: A utilitarian benefits vs harms analysis regarding research participants. Under this branch researchers are expected to maximize benefits to research participants, to identify and minimize harms, and determine mitigation strategies for harms that remain.
\item \emph{Justice and Fairness}: Researchers should try to consider and distribute the benefits and burdens of the research equally. This can apply to participant selection, the study itself, and the outputs and goals of the study.
\item \emph{Law and Public Interest}: While the Menlo report makes clear that researchers are not expected to be legal experts, it does ask researchers to do their due diligence in trying to ensure their study is fully legal, to be as transparent as possible, and to hold themselves accountable if something goes wrong. Additionally, the study should hold some public value.
\end{enumerate}

While the Menlo report is not a fully comprehensive ethical framework (and in particular ignores several difficult questions of computer security ethics including how to responsibly manage vulnerability disclosures \cite{ISOIEC29147,householderCERTGuideCoordinated,FosteringResponsibleComputing})
we find this to be a good place to start when considering standard informed consent practices and times when the use of deception may be justified.
In many studies, these principles are applied in a relatively straightforward way.
In the best case scenario, researchers gather informed consent from all involved in the study \cite{menloReport,belmontReport,InformedConsentFAQs,cranorConferenceSubmissionReview},
much computer security research involves minimal or no harms, and, with proper participant selection and study design, raises few or no issues of justice and fairness.

Ethical restrictions always require researchers to protect the privacy of those involved with their study \cite{menloReport,ethicsOfOnlineResearch,alllmanIssuesEtiquetteConcerning2007}.
Researchers should also take into account whether users would generally expect collected information to be private or not, especially in the context of social media \cite{ethicsOfOnlineResearch}.
This can pose especially thorny issues when using a public dataset that was the result of a leak \cite{thomasEthicalIssuesResearch2017}.
As identified in the Menlo report, computer security research often needs ``protection of human subjects'' to apply not only to study participants, but also to other humans in the system, and occasionally to computer systems as well.
Taking as an example a hypothetical web scraping study, a typical study design would not have users whose data was scraped sign a consent form.
However,  researchers are expected to protect the interests of the individuals whose data is represented and are expected to avoid taking illegitimate action to obtain datasets \cite{menloReport}. (However, the dataset may still be of illicit origin even if the researchers did nothing illegitimate, for example if the dataset was leaked and made publicly available.) These issues are not directly related to deception and are addressed in other ethical analyses of computer science research, e.g.\ \cite{thomasEthicalIssuesResearch2017,partridgeEthicalConsiderationsNetwork2016,menloReport,cranorConferenceSubmissionReview}.

\textbf{The ethics of deception.}
In the context of this work on fraud, when computer science research involves deception on the part of the researchers or their agents -- whether that deception is directly to a person, to a computer system, or involves ``bystanders'' in the wild -- that poses a major ethical hurdle.

As is true in other research fields like social or biomedical research, the use of deception should not be taken lightly and raises an ethical barrier that the study must overcome if it is justified (and indeed some would argue that deception is never justified in research under any circumstances \cite{baumrindResearchUsingIntentional1985,meadResearchHumanBeings}, although many ethical frameworks in research permit deception under some circumstances \cite{findleyObligatedDeceiveAliases2016,goodeEthicsDeceptionSocial1996,InformedConsentFAQs,belmontReport,wendlerDeceptionPursuitScience2004}).

That said,
we see deception in computer security as  different from deception in social and biomedical research.  Tying in with our discussion of deception in \Cref{sec:def-deception}, this is because:
\begin{enumerate}
\item The deception is often \emph{to a computer system} rather than a person
\item The deception is often done \emph{by a computer system} rather than a person
\item The purpose of the deception is often to \emph{test the security or boundaries} of a system, and therefore often violates written or unwritten rules of the system
\end{enumerate}

This suggests the need to adapt ethical guidelines for dealing with deception into a computer security context.

\subsection{Justified uses of deception in computer security research}

Based upon 
the Common Rule's criteria for a waiver of informed consent \cite{InformedConsentFAQs},
the Menlo and Belmont reports \cite{menloReport,belmontReport},
ethics considerations in computer security and measurement \cite{FosteringResponsibleComputing,partridgeEthicalConsiderationsNetwork2016,kohnoEthicalFrameworksComputer,cranorConferenceSubmissionReview},
the history of laws governing ethical restrictions on experiments \cite{curranGovernmentalRegulationUse1969,finnEthicsGovernanceDevelopment2023},
justifications for deception in social and other research \cite{goodeEthicsDeceptionSocial1996,christensenDeceptionPsychologicalResearch1988,findleyObligatedDeceiveAliases2016},
and post-mortems of prominent ethical controversies in computer security \cite{narayananNoEncoreEncore2015,jakobssonWhyHowPerform2008},
we suggest that the use of deception in computer security can be justified when the following conditions are met:

\textbf{1. The ``least deceptive means'' are used; less deceptive methods are not feasible or effective.}
Given that concerns about deception in research largely derive from \emph{respect for persons} \cite{menloReport,belmontReport,FosteringResponsibleComputing} (and, to some extent, concerns about maintaining trust in science \cite{baumrindResearchUsingIntentional1985,meadResearchHumanBeings,robertt.bowerEthicsSocialResearch1978,curranGovernmentalRegulationUse1969}),
we believe that deception can be justified if researchers are respecting persons as much as the study goals reasonably allow.
Study design plays a big part in this ethical determination, just as it does in measurement studies \cite{partridgeEthicalConsiderationsNetwork2016}.
Controlled lab experiments rather than experiments ``in the wild'' can remove this element of deception, when feasible.
Additionally, in keeping with the respect for persons principle and keeping in mind the blurrier line between ``hacking'' and ``tricking'' machines \cite{calo2018}, we see deception \emph{of systems and machines} as less directly problematic than deception \emph{of people} -- with the caveat that indirect deception of people is still deception of people.

\textbf{2. The study's value warrants the use of deception.}
From a more consequentialist perspective, most computer security studies are intended to measure or improve the security of systems used by many people, and thus provide a justification for performing the research.  The value of the knowledge gained as a result of the study is a necessary but not sufficient condition for the study to be ethical.  In particular, a study that provides minimal value even if the study went perfectly well may not be sufficient justification to overcome the ethical issues arising from deception.  This principle is also found in justifications of deception in psychology research \cite{christensenDeceptionPsychologicalResearch1988,goodeEthicsDeceptionSocial1996,findleyObligatedDeceiveAliases2016}.

\textbf{3. ``Bystanders'' involved with the deceived system will face minimal or no harms, rights violations, or injustices as a result of the deception.}
Inspired by the Common Rule's requirements for a waiver of informed consent \cite{InformedConsentFAQs} and similar concerns about minimizing harms in measurement studies \cite{partridgeEthicalConsiderationsNetwork2016}, we believe that no amount of deception would be justified in a study if ``bystanding'' users were to be significantly harmed as a result.
This principle also underlies existing ethical principles in computer security for ensuring that one does not hack ``live'' systems that will have an impact on users \cite{kohnoEthicalFrameworksComputer} and ensure the privacy and confidentiality of ``bystander'' users is protected \cite{menloReport,alllmanIssuesEtiquetteConcerning2007,cranorConferenceSubmissionReview}.

\begin{acks}

The authors would like to thank Boston University School of Law students Madeleine Bomberg and Will Izenson from the BU/MIT Student Innovations Law Clinic for their legal assistance in the research described in the article.

Author Madelyne Xiao was supported by NSF DGE-2039656.

Author Andrew Sellars is a partner at the law firm Albert Sellars LLP, which did not contribute any funds toward the drafting of this article. The views in this paper may not represent the views of the firm or any of its clients.

\end{acks}

\appendix
\section{Summary table showing our taxonomy of computer security research}
\label{sec:table}
Table \ref{tab:research_taxonomy} contains a table which summarizes our taxonomy of computer security research methods that could be considered deceptive, from \Cref{sec:taxonomy}.

\begin{table*}[ht]
    \def\arraystretch{1.4}
    \footnotesize
    \centering
    \caption{A summary of our taxonomy of computer security research methods that could be considered deception (see \Cref{sec:taxonomy})}
        \begin{tabular}{p{2cm}|p{6.0cm}p{6.0cm}p{2cm}} 
        Research method & Method description & Uses of deception in research & Exemplar(s) \\
        \midrule \midrule
        (\ref{sec:pentesting}) Penetration testing \& red teaming  & Hacking software systems to identify and report vulnerabilities in the target system. These functions are often performed by whitehat hackers acting independently or on behalf of the target entity (e.g., a government agency or company). & Researchers spoof network information or incorrect credentials. \cite{gallowayPracticalAttacksDNS2024} game the DNS reputation system to artificially increase the reputation score of a suspicious domain.  & \cite{simons2001, gallowayPracticalAttacksDNS2024, maFakeBehalfImperceptibleEmail} \\
        \hline
        (\ref{sec:passive_mmt}) Web scraping, snapshotting, measurement & Conducting automated or manual collection of publicly visible web content. & Researchers perform passive measurement or observation of online communities; they may provide only partial disclosure of their identities, may refrain from divulging their presence entirely, or may redact or spoof network information in their interactions. \cite{decary2014policing} use bots to infiltrate IRC chat rooms and scrape chat logs without disclosing the presence of these bots to chatroom participants. & \cite{decary2014policing, kreibichSpamCampaignTrail2008} \\
        
        \hline
        (\ref{sec:scripted}) Scripted user studies & Recruiting study participants perform a pre-specified series of actions using their own identities and/or account credentials. This method---and its attendant risks to the surrogate---is often preferable to (e.g.) sock puppets for reasons of ecological validity. & Researchers instruct study surrogates to misrepresent themselves or their intentions to a target system. \cite{haeder2016secret} instruct ``secret shoppers'' to request healthcare provider information from insurance marketplaces. & \cite{haeder2016secret, dimkov2010two} \\
        \hline
        (\ref{sec:sock_puppets}) Sock puppets & Creating and operating fake accounts. Though deception is a core element of this work---the accounts are not operated by the person(s) whose name(s) they bear---the \textit{degree} of deception varies with the parties (e.g., a platform algorithm, other users) most likely to interact with these fake accounts. & Researchers generate fake accounts, oftentimes from multiple IP addresses or devices, in order to circumvent bot detection mechanisms; they might use these accounts to engage with real (human) users who are unaware that they are interacting with a sock puppet. \cite{boshmaf2011socialbot} infiltrate Facebook social graphs with a botnet comprising about a hundred bot accounts. & \cite{boshmaf2011socialbot, srba2023auditing, bandy2021more} \\
        \hline
        (\ref{sec:soc_eng}) Social engineering and Interactive Studies & Interacting with real human users to measure user or system responses to perturbations. & Researchers send false or misrepresentative email communications via spam or phishing campaigns, or perturb online platforms to observe user and platform responses. \cite{royChatbotsPhishbotsPhishing2024} deploy LLM-generated phishing attacks. & \cite{di2022revisiting, acharyaConningCryptoConman2024, jakobssonWhyHowPerform2008, royChatbotsPhishbotsPhishing2024} 
        \\
        \hline
        (\ref{sec:audits}) Audits & Understanding algorithm functionality under-the-hood via selective manipulation of inputs and observation of corresponding outputs.  & Researchers submit false or misrepresentative data points to a system in order to observe its outputs. \cite{kaplanMeasurementAnalysisImplied2022} posed as Facebook advertisers, submitted multiple job postings, and observed potential biases in the distribution of these ads. & \cite{kaplanMeasurementAnalysisImplied2022, westPictureWorth5002024, chengALIFLowCostAdversarial2024, jiangCanHearYour, kimScoresTellEverything2024} \\
        \hline
        (\ref{sec:adversarial_ml}) Attacks on ML & Developing attacks on ML systems in order to test their robustness, infer training set contents or guardrails, or poison training data. & Researchers submit false or misrepresentative prompts to a system and observe its outputs. \cite{perez2022red} use a language model to generate adversarial attacks on an LLM. & \cite{perez2022red, yanLLMAssistedEasyTriggerBackdoor, liuAFGenWholeFunctionFuzzing2024, yangSneakyPromptJailbreakingTextimage2024, yuDontListenMe2024} \\ 
        \hline \hline
        (\ref{sec:id_work}) ID Research & Research involving presentation of real or prop ID documentation to an AI/ML-driven verification system. Prop IDs might be modified forms of genuine ID documents. & Researchers submit false or misrepresentative images to a black box system and observe its outputs. \cite{di2019personal} modify photocopies of ID images and observe the efficacy of ID verification systems on these modified documents.  & \cite{di2022revisiting, di2019personal, zhao2021deep, engelbertzSecurityAnalysisEIDAS} 
        \\
        \hline \hline

        \end{tabular}
\label{tab:research_taxonomy}
\end{table*}
\section{Long-discussed legal concerns in computer security research}
\label{sec:cfaa-dmca}

Research that engages in deception must of course also confront the longstanding issues that all software and platform researchers face under laws like the CFAA, DMCA, and claims rooted in breaches of terms of service. While there remain many aspects of these laws that still present legal risk, the risks have worked to become at least more well understood, and in some respects considerably improved. This subsection reviews the current understanding within these laws.

\subsection{CFAA}
The CFAA has long been an obstacle for those seeking to do software and platform research \cite{parkResearchersGuideLegal, baranetsky2018}. The law contains close to a dozen independent theories of liability, all circling around the notion that to be actionable one must access a computer or transmit information to a computer either "without authorization," or by "exceeding authorized access." (18 U.S.C. §1030.) There is no research exemption to the CFAA, \cite{calo2018} although experts have called for a safe harbor that would encompass this and other laws \cite{abdo_safe_2022, longpre_safe_2024}.

Nearly all of the techniques discussed above have had to consider risk of civil or criminal liability under the CFAA, with courts struggling for well over a decade to apply the law to common information gathering techniques such as web scraping \cite{sellars2018}. But the scope of the CFAA was helpfully narrowed in 2021 in the Supreme Court's decision \textit{Van Buren v. United States}. In that case, the Supreme Court addressed how to interpret when one "exceeds authorized access," and in particular whether one violates the law when one \textit{has} access to a computer, but uses it for an \textit{impermissible purpose}. The court said no; it did not go as far as to say it would only consider technical restrictions on access to computers, but made clear that those who have at least some access to computer systems do not face CFAA liability for using their access in ways the computer owner dislikes.\footnote{Van Buren v. United States, 593 U.S. 374, 391–92 (2021).} 

The approach taken in \textit{Van Buren} has eased, but not removed, concerns about research colliding with the CFAA \cite{bhandari2024}. As summarized by Park and Albert, "research that only accesses your own devices does not violate the CFAA," and "[i]f you are working with devices that you don't own, but all resources that you access on those devices are either publicly available or you have valid credentials (issued to you) that provide access to them, there is no CFAA violation" \cite{parkResearchersGuideLegal}. The law is still present in experiments rooted in penetration testing and red teaming, and application of the CFAA to adversarial attacks on AI remains a significant open question \cite{albert_ignore_2024}. But other forms of research have seen their criminal risk mitigated substantially, even while other legal theories like breach of contract claims or trespass to chattel have begun to fill in this gap \cite{sobel2021}.

And even where there remains CFAA risk, the risk of federal prosecution has apparently eased as well. The Department of Justice in 2022 updated its Justice Manual to instruct prosecutors to "decline prosecution if available evidence shows the defendant's conduct consisted of, and the defendant intended, good-faith security research," borrowing its definition from the most recent triennial DMCA exemption rulemaking, discussed in the next section \cite{OfficePublicAffairs2022}. It is not clear that this represents a practical change in enforcement strategy \cite{pfefferkorn2022}, but it can be used as an effective legitimizing signal for this form of work.\footnote{Indeed, the fact that the prior version of this policy \textit{was not} binding allowed a prior constitutional challenge to the CFAA proceed. \textit{See} Sandvig v. Sessions, 315 F. Supp. 3d 1, 20 (D.D.C. 2018).} 

Finally, some courts have cut off a threat for civil action under the CFAA by picking up on dicta in the \textit{Van Buren} decision that purports to limit "loss" to "technological harms" only,\footnote{\textit{Van Buren}, 593 U.S. at 376.} and using that deny civil claims when the alleged intrusion only resulted in costs related to investigation of the incident and consulting with technology lawyers.\footnote{\textit{See, e.g.}, X Corp. v. Ctr. for Countering Digital Hate, Inc., 724 F. Supp. 3d 948, 983 (N.D. Cal. 2024), \textit{appeal docketed}, No. 24-2643 (9th Cir. April 25, 2024) (denying CFAA "loss" for attempting to conduct investigations into the extent of unauthorized access and legal expenses).} This approach has not been universal -- \cite{pfefferkorn2022} illustrates an emerging split between courts in how to internalize this indirect instruction from \textit{Van Buren} -- but a move in that direction would provide some of the same practical risk mitigation discussed in relation to fraud above.

\subsection{DMCA} 
Another common legal concern for platform research stems from the "anticircumvention" provisions which were introduced to United States copyright law in the Digital Millennium Copyright Act, enacted in the late 1990s to address concerns about digital piracy. The provisions prohibit the "circumvention" of a "technical protection measure" that controls access to a copyright-protected work. Because software programs and websites are typically protected under copyright, any bypassing of a technical hurdle to access a work at least arguably implicates the law. Legal risk under the DMCA most frequently arises when the research involves bypassing controls like encryption, CAPTCHAs, authentication handshakes, and other technical gates. Some forms of auditing and scraping, as well as attacks on machine learning systems and penetration testing, tend to be the areas of greatest risk under the DMCA.\footnote{\textit{See, e.g.}, Yout, LLC v. RIAA, 633 F. Supp. 3d 650 (D. Conn. 2022), \textit{appeal docketed}, No. 22-2760 (2d Cir. argued Feb. 5, 2024) (finding that bypassing YouTube's technical provisions against downloading videos was likely a DMCA violation).}

Here, too, the law has gradually moved into a more tolerable state for research, albeit with many remaining concerns. As reviewed by Park and Albert, there are both permanent exemptions written into the statute and expanded temporary (but frequently renewed) exemptions for a broader set of research activity that may require the researcher to bypass a technical protection measure \cite{parkResearchersGuideLegal}. There is also a good argument that much of the work of bias research does not implicate the DMCA, as it does not require the researcher to bypass any technical measures. Web scraping, creation of sock puppet accounts, audits, and scripted user studies that use the platforms as designed and intended similarly do not require a researcher to bypass a technical measure, unless there are technical safeguards in place to specifically control that practice (by, for example, presenting a CAPTCHA after a certain number of accounts are created or pages are accessed). \textit{Tooling} for research remains a significant issue, as the regulatory exemptions of the DMCA apply only to individual acts of circumvention, and not the distribution of tools that are used to facilitate such circumvention, which are separately prohibited in 17 U.S.C. §§1201(a)(2), 1201(b)(1) \cite{parkResearchersGuideLegal}.

\subsection{Terms of Service.} 
\label{sec:tos}
Courts that narrowed the scope of the CFAA to activities like web scraping were quick to observe that they were still leaving platform owners recourse in other areas, including lawsuits rooted in the website's terms of service.\footnote{\textit{See} hiQ Labs, Inc. v. LinkedIn Corp., 31 F.4th 1180, 1201 (9th Cir. 2022) ("Entities that view themselves as victims of data scraping are not without resort, even if the CFAA does not apply: state law trespass to chattels claims may still be available. […] And other causes of action, such as copyright infringement, misappropriation, unjust enrichment, conversion, breach of contract, or breach of privacy, may also lie.").} And indeed, present fights around website access gravitate around these contracts---despite the near-unanimous view that these contracts are rarely read or understood and often contain ambiguous or inconsistent language \cite{fiesler_no_2020}. Terms of Service are drafted to press every advantage for the online platform, and routinely prohibit the sorts of activity essential for platform research \cite{bhandari2024}. The past few years have seen an uptick in this form of enforcement, ranging from a platform directly suing a critic for alleged violations on its terms against web scraping,\footnote{\textit{Ctr. for Countering Digital Hate}, 724 F. Supp. 3d 1.} to use of terms of service to justify suspension of accounts, perhaps most famously the accounts of Laura Edelson and others at New York University who gathered data that examined Meta's handling of political advertising in 2020 and 2021 \cite{longpre_safe_2024, gilbert2024risks}. 

Here, too, researchers and others who act in ways in tension with platform terms have begun some effective lines of defense, inspired by cautions by courts about application of the law in ways that "risk[] the possible creation of information monopolies that would disserve the public interest."\footnote{\textit{hiQ Labs}, 31 F.4th at 1202.} This can include very technical and careful reading of terms and more advanced argumentation around when those terms are binding on the user,\footnote{\textit{See, e.g.}, Meta Platforms, Inc. v. Bright Data Ltd., No. 23-cv-77, 2024 WL 251406 at **10, 15 (N.D. Cal. Jan. 23, 2024) (because defendant was not acting as a "user" of Facebook at the time of their scraping of publicly-available information, the Facebook terms of service did not apply to the activity); \textit{but see} X Corp. v. Bright Data Ltd., 733 F. Supp. 3d 832, 847 (N.D. Cal. 2024) (because X's terms of service purported to reach nonusers as well and defendant was aware of that, agreement was validly formed)} arguments that confine the recoverable damages related to the breach to zero or near enough to zero to be a tolerable risk,\footnote{\textit{Ctr. for Countering Digital Hate}, 724 F. Supp. 3d at 969–70 (finding that X could recover neither direct nor special damages from an alleged breach by a research organization); \textit{Bright Data}, 733 F. Supp. 3d at 847–48. Similar approaches have been used to limit claims under the theory of "trespass to chattel," which has become a popular theory by which to pursue web scraping defendants given the new confines of the CFAA. \textit{See Bright Data}, 733 F. Supp. 3d at 843.} and arguments that certain forms of terms of service violations are preempted by copyright law.\footnote{\textit{Bright Data}, 733 F. Supp. 3d at 852–53; Genius Media Group Inc. v. Google LLC, No. 19-cv-7279, 2020 WL 5553639 (E.D.N.Y. Aug. 10, 2020), \textit{aff'd sub nom.} ML Genius Holdings LLC v. Google, No. 20-3113 (2d Cir. March 10, 2022) (unpublished).} The Federal Trade Commission has also brought actions under the Consumer Review Fairness Act (15 U.S.C. §45b) against companies that attempt to use a terms of service to stifle consumer assessments and reviews \cite{ensor2024}. What remains an issue, however, is how to respond to platforms that enforce terms through technological self-help, such as blocking users or IP addresses. Because there is little \textit{right} to be on a given platform, response to such suspension of accounts can be especially challenging \cite{longpre_safe_2024}.

\bibliographystyle{ACM-Reference-Format}
\bibliography{references}

\end{document}